\documentclass[
    reprint,
    longbibliography,
    superscriptaddress,
    floatfix,
    amsmath,
    amssymb,
    aps,
    prd
]{revtex4-2}

\usepackage[T1]{fontenc}
\usepackage[utf8]{inputenc}

\usepackage{xcolor}
\usepackage{graphicx}

\definecolor{APSLinkBlue}{RGB}{46,48,146}
\usepackage{hyperref}
\definecolor{LinkBlue}{RGB}{0,0,255}
\hypersetup{
  colorlinks=true,
  linkcolor=blue,
  citecolor=blue,
  urlcolor=blue,
  filecolor=blue
}

\hypersetup{pdfsubject={DOI: 10.1103/mqsp-mwbs}}
\setcitestyle{citesep={,}}
\begin{document}

\title{Thermodynamics and mechanical laws of causal-diamond horizons}

\author{Carlos R. Ordóñez}
\affiliation{Department of Physics, University of Houston, Houston, Texas 77204, USA}
\author{Gustavo Valdivia-Mera}
\email{Contact author: gvaldiviamera@uh.edu}
\affiliation{Department of Physics, University of Houston, Houston, Texas 77204, USA}
\affiliation{Texas Center for Superconductivity, University of Houston, Houston, Texas 77204, USA}
\makeatletter\def\Dated@name{}\makeatother
\date{Received 11 August 2026; accepted 9 September 2026}

\begin{abstract}
We develop a unified thermodynamic and mechanical framework for causal-diamond horizons in Minkowski spacetime. Euclidean regularity fixes the temperature $T_D=1/(\pi\alpha)$, where $2\alpha$ is the temporal extent of the diamond. The regulated Gibbons-Hawking-York contribution gives $I_D^E=-A_0/(4G)$, with $A_0$ the area of the horizon bifurcation surface, and the vanishing equilibrium geometric energy then yields the entropy-area relation $S_D=A_0/(4G)$. For weak matter perturbations crossing a localized portion of the future horizon, a Raychaudhuri analysis relates the associated Killing-energy flux to the area response, yielding a local physical-process first law at leading order near the bifurcation surface and, for classical matter satisfying the null energy condition, a local second law. Uniform conformal surface gravity establishes the zeroth law, while zero temperature is approached only in the limit $\alpha\to\infty$, in which the finite causal horizon disappears. These results show that a coherent horizon-law structure can arise within a causal diamond, a fundamental building block of the causal structure of Minkowski spacetime. The derivation is explicit in $(2+1)$ dimensions, while the physical results retain the same form for spherical causal diamonds in $d\geq3$.
\end{abstract}

\maketitle
\newpage
\tableofcontents

\section{Introduction}
\label{sec:introduction}

The thermodynamic interpretation of horizons emerged from a sequence
of results connecting classical geometry with temperature and
entropy. Bardeen, Carter, and Hawking organized the properties of
stationary black holes into four mechanical laws
\cite{BardeenCarterHawking1973}. Bekenstein subsequently associated
black-hole entropy with horizon area \cite{Bekenstein1973}, while
Hawking's particle-creation calculation fixed the corresponding
temperature \cite{Hawking1975}. The Euclidean formulation of Gibbons
and Hawking then provided a geometric framework in which regularity
determines the temperature and the gravitational action encodes
the horizon entropy \cite{GibbonsHawking1977,hawking1979path}.

These ideas extend beyond black-hole event horizons. The Unruh effect
associates a temperature with the Rindler horizon of a uniformly
accelerated trajectory in Minkowski spacetime \cite{Unruh1976}, and
Laflamme showed that Euclidean Rindler geometry yields the
corresponding entropy-area contribution directly from the
gravitational action \cite{Laflamme1987}. Jacobson later demonstrated that the Einstein equation follows from
imposing a local Clausius relation together with an entropy-area
proportionality for local Rindler horizons \cite{Jacobson1995}.
Likewise, for quasistationary perturbations, the physical-process
first law relates stress-energy flux through the horizon to its area
variation \cite{AmselMarolfVirmani2008}. These developments indicate that the
thermodynamic properties of horizons are closely tied to their local
causal and geometric structure.

Causal diamonds provide a finite setting in which this connection can
be examined. Their null boundaries are preserved by a distinguished
conformal Killing flow, whose associated temperature was identified
by Martinetti and Rovelli \cite{martinetti2003diamond}. For conformal
field theories, the conformal map from a spherical region to a
hyperbolic geometry makes the corresponding thermal structure
explicit \cite{casini2011towards}. Complementary analyses have
examined causal-diamond thermality through open-system dynamics
\cite{chakraborty2022thermal}, conformal quantum mechanics
\cite{arzano2020conformal,camblong2024conformal}, a path-integral
construction of the thermofield-double state
\cite{chakraborty2025path}, and the analytic structure of the
inertial propagator \cite{eissa2025thermal}. The quantum-information
consequences of the same causal restriction have also been explored
through entanglement degradation induced by the restricted access of
finite-lifetime diamond observers
\cite{camblong2024entanglement}. Other studies have addressed the
geometric thermodynamic structure of causal diamonds, including a
Smarr relation and a variational first law
\cite{jacobson2019gravitational}, an entropy contribution from the
Euclidean boundary term surrounding the causal-diamond puncture
\cite{banks2021path}, and a replica analysis of modular-Hamiltonian
fluctuations \cite{FransenHeZurek2025}. Separately, related null
geometries have been studied through area-comparison and rigidity
results for light-cone cross sections
\cite{ChoquetBruhatChruscielMartinGarcia2009}, and through analogues
of the four laws of black-hole thermodynamics for bifurcate conformal
Killing horizons extending to null infinity
\cite{DeLorenzoPerez2018}.

This literature motivates a more specific question: whether the
equilibrium thermodynamic properties of a finite causal-diamond
horizon and its response to matter perturbations can be incorporated
within a single horizon-mechanics framework. We address this question
for causal diamonds in Minkowski spacetime by treating the Euclidean
and Lorentzian descriptions as complementary aspects of the same
finite causal geometry. The former characterizes the equilibrium
horizon, whereas the latter determines its local response away from
equilibrium. This construction provides a unified setting in which
the thermodynamic and mechanical properties of the causal-diamond
horizon can be studied together.

This paper is organized as follows. In
Sec.~\ref{sec:equilibrium-thermodynamics}, we introduce the
causal-diamond geometry and its diamond-time conformal Killing flow
and develop the corresponding equilibrium thermodynamics from the
Euclidean description. In Sec.~\ref{sec:mechanical-laws}, we develop
the mechanical laws of the causal-diamond horizon by considering both
its local response to a weak stress-energy perturbation and the
equilibrium geometry of the complete unperturbed diamond. For
geometric transparency, the analysis is carried out in
\((2+1)\)-dimensional Minkowski spacetime, while the extension to
spherical causal diamonds in \(d\geq3\) is presented in
Appendix. The
\(d=2\) case lies outside the scope of the present gravitational
construction, since the Einstein tensor vanishes identically and the
zero-dimensional horizon cross sections do not admit the
transverse-area variation used in our analysis. Nevertheless, aspects
of causal-diamond thermodynamics in $(1+1)$-dimensional spacetime have been explored
elsewhere \cite{JacobsonVisser2023Ensembles}. We conclude by
summarizing the resulting thermodynamic and mechanical framework and
discussing its physical implications.

Throughout this work, we use natural units
\(c=\hbar=k_B=1\).

\section{Euclidean path integral and equilibrium thermodynamics of the causal diamond}
\label{sec:equilibrium-thermodynamics}

The equilibrium thermodynamic description of the causal diamond will
be developed from the Euclidean gravitational path integral
\cite{GibbonsHawking1977,hawking1979path},
\begin{equation}
Z_{\rm grav}(\beta)
=
\int_{\mathcal G_\beta}
\mathcal Dg\,
e^{-I_E[g]}
\simeq
e^{-I_E[g_\beta]},
\label{eq:gravitational-path-integral}
\end{equation}
where \(\mathcal G_\beta\) denotes the class of Euclidean
configurations compatible with Euclidean period \(\beta\) and the
prescribed boundary data, \(I_E[g]\) is the Euclidean gravitational
action, and \(g_\beta\)
denotes the Euclidean configuration entering the semiclassical
approximation.

In \((2+1)\) dimensions, the Euclidean Einstein action, including the
Gibbons-Hawking-York boundary term
\cite{GibbonsHawking1977,York1972}, is
\begin{equation}
\begin{aligned}
I_E[g]
&=-\frac{1}{16\pi G}\int_{\mathcal M_E}d^3x\,\sqrt g\,(R-2\Lambda)\\
&\quad-\frac{1}{8\pi G}\int_{\partial\mathcal M_E}d^2x\,\sqrt h\,K.
\end{aligned}
\label{eq:Euclidean-gravity-action}
\end{equation}
Here, \(\mathcal M_E\) is a Euclidean manifold and
\(\partial\mathcal M_E\) is its boundary. Moreover,
\(g=\det g_{\mu\nu}\), \(R\) is the Euclidean Ricci scalar,
\(\Lambda\) is the cosmological constant, and \(G\) is Newton's
constant. On \(\partial\mathcal M_E\), \(h_{ab}\) is the induced
metric, \(h=\det h_{ab}\), and \(K\) is the trace of its extrinsic
curvature
\cite{York1972,GibbonsHawking1977,hawking1996gravitational}.

For the unperturbed causal diamond considered here, the underlying
Lorentzian manifold is Minkowski spacetime and \(\Lambda=0\). Its regular
Euclidean continuation is therefore flat, with \(R=0\), so the
complete regular bulk contribution in
Eq.~\eqref{eq:Euclidean-gravity-action} vanishes on shell.
Consequently, the nontrivial contribution associated with the
causal-diamond horizon is carried by the gravitational boundary term.

As the following geometric construction will show, the causal
horizons are null in the underlying Lorentzian geometry and are
reached only as asymptotic limits of the causal-diamond chart. For
precisely this reason, the horizon contribution must be obtained as a
regulated limit from within the diamond. To implement this limit, we
follow the standard stretched-horizon prescription
\cite{PriceThorne1986,ParikhWilczek1998}: a nearby timelike
regulating hypersurface is introduced inside the Lorentzian diamond,
its Euclidean continuation provides the smooth boundary on which the
boundary term is evaluated, and the horizon limit is taken only
afterward~\cite{Laflamme1987}. This construction will allow us to
evaluate the horizon contribution and thereby extract the
causal-diamond thermodynamic relations.

The construction in this section is carried out explicitly in
\((2+1)\) spacetime dimensions. Its extension to spherical causal
diamonds in general \(d\geq3\) spacetime dimensions is developed in
Appendix.

\subsection{Causal-diamond geometry and conformal flow}
\label{subsec:causal-diamond-geometry}

Consider a rotationally symmetric causal diamond \(D\) in
\((2+1)\)-dimensional Minkowski spacetime \(\mathbb{R}^{2,1}\),
whose central timelike line
\(r=0\) extends over the interval \(-\alpha<t<\alpha\), of total
duration \(2\alpha\), with spatial polar coordinates
\((r,\varphi)\):
\begin{equation}
D
=
\left\{
(t,r,\varphi): |t|+r<\alpha
\right\}.
\label{eq:causal-diamond-domain}
\end{equation}
Hereafter, we refer to \(D\) simply as the diamond.

\begin{figure}[t]
\centering
\includegraphics[width=0.82\columnwidth]{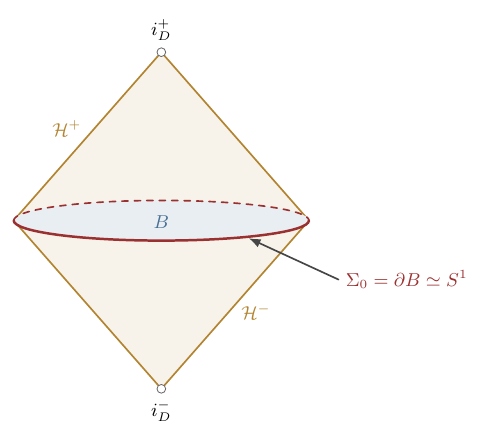}
\caption{
Schematic representation of a rotationally symmetric causal diamond in
\((2+1)\)-dimensional Minkowski spacetime. The horizons
\(\mathcal H^\pm\) meet at the waist
\(\Sigma_0=\partial B\) and terminate at the tips
\(i_D^\pm\).
}
\label{fig:causal-diamond-2p1}
\end{figure}

The diamond is equivalently defined as the domain of dependence of the
central spatial disk
\begin{equation}
B
=
\left\{
(t,r,\varphi):
t=0,\quad r<\alpha
\right\}.
\label{eq:central-cauchy-ball}
\end{equation}
As illustrated in Fig.~\ref{fig:causal-diamond-2p1}, the upper half
of \(D\) is the future domain of dependence of \(B\): every
past-inextendible causal curve through a point in this region
intersects \(B\). Likewise, the lower half is its past domain
of dependence, since every future-inextendible causal curve through
a point in this region intersects \(B\). Thus, \(B\) is a Cauchy surface for \(D\) and its maximal spatial
section. Its boundary is the waist,
\begin{equation}
\Sigma_0
=
\partial B
\simeq S^1,
\label{eq:waist-as-boundary-of-cauchy-ball}
\end{equation}
which, as illustrated in Fig.~\ref{fig:causal-diamond-2p1}, is the
bifurcation surface where the future and past horizons
\(\mathcal H^\pm\) meet. Throughout this discussion, codimension is
measured relative to the full spacetime. Hence, \(\Sigma_0\) has
codimension two.

For the unperturbed diamond, the induced line element on the waist is
\(ds_{\Sigma_0}^2=\alpha^2d\varphi^2\). The corresponding waist area,
given by the circumference of \(\Sigma_0\), is
\begin{equation}
A_0
=
\int_0^{2\pi}
\alpha d\varphi\,
=
2\pi\alpha.
\label{eq:causal-diamond-area}
\end{equation}

The future and past horizons have the radial representations
\begin{align}
\mathcal H^+
&:\quad
t+r=\alpha,
\qquad
0<t<\alpha,
\label{eq:CD-null-boundaries-plus}
\\
\mathcal H^-
&:\quad
-t+r=\alpha,
\qquad
-\alpha<t<0,
\label{eq:CD-null-boundaries-minus}
\end{align}
while the tips are located at
\((t,r)=(\pm\alpha,0)\). Away from the tips, the spatial sections of
the horizons are circles.

To describe the geometry in coordinates adapted to the causal
diamond, we introduce the chart
\((\eta,\rho,\varphi)\) on the interior of \(D\). Its relation to
the Minkowski coordinates is
\cite{casini2011towards,arzano2020conformal}
\begin{align}
t(\eta,\rho)
&=
\alpha
\frac{
\sinh(2\eta/\alpha)
}{
\cosh\rho+\cosh(2\eta/\alpha)
},
\label{eq:CD-conformal-map-t}
\\
r(\eta,\rho)
&=
\alpha
\frac{
\sinh\rho
}{
\cosh\rho+\cosh(2\eta/\alpha)
}.
\label{eq:CD-conformal-map-r}
\end{align}
The ranges
\(-\infty<\eta<\infty\),
\(0\leq\rho<\infty\), and
\(0\leq\varphi<2\pi\) cover the interior of the diamond. Its causal
boundary is not part of the chart and is reached only through
asymptotic limits of \(\eta\) and \(\rho\).

\begin{figure*}[t]
\centering
\begin{minipage}{0.42\textwidth}
\centering
\includegraphics[width=\linewidth]{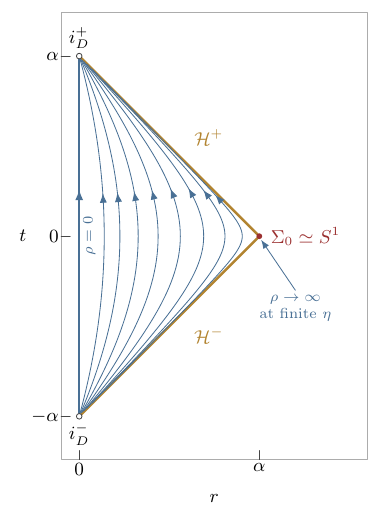}\par
\smallskip
(a)
\end{minipage}
\hspace{0.6cm}
\begin{minipage}{0.42\textwidth}
\centering
\includegraphics[width=\linewidth]{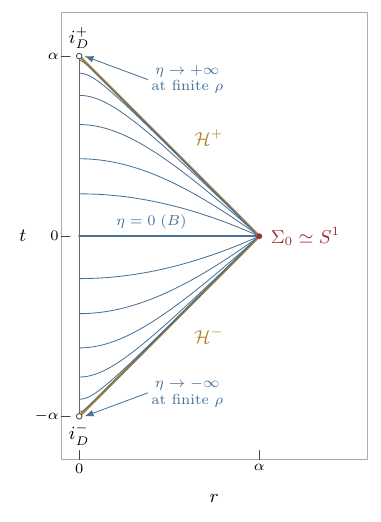}\par
\smallskip
(b)
\end{minipage}
\caption{
Analytic coordinate curves in the radial plane of the causal diamond. (a) Curves of constant \(\rho\), including the central timelike
trajectory \(\rho=0\).
(b) Slices of constant \(\eta\), with \(\eta=0\) corresponding to
the central Cauchy disk \(B\).
}
\label{fig:CD-radial}
\end{figure*}

Several geometric features follow directly from
Eqs.~\eqref{eq:CD-conformal-map-t} and
\eqref{eq:CD-conformal-map-r}, as illustrated in
Fig.~\ref{fig:CD-radial}. The curve \(\rho=0\) coincides with the
central timelike trajectory \(r=0\), while the slice \(\eta=0\)
coincides with the central Cauchy disk \(B\). Curves of
constant \(\rho\) run from the past tip to the future tip, whereas
slices of constant \(\eta\) provide a spacelike foliation of the
diamond. At fixed finite \(\eta\), the limit
\(\rho\to\infty\) approaches the waist \(\Sigma_0\), while the
future and past tips are approached as
\(\eta\to\pm\infty\) at fixed \(\rho\). Generic points on the
future and past horizons require correlated asymptotic limits of
\(\eta\) and \(\rho\).

Substituting Eqs.~\eqref{eq:CD-conformal-map-t} and
\eqref{eq:CD-conformal-map-r} into the Minkowski line element, $ds^2=-dt^2+dr^2+r^2d\varphi^2$, gives
\begin{equation}
ds^2
=
\Omega^2(\eta,\rho)
\left[
-4\,d\eta^2
+
\alpha^2d\rho^2
+
\alpha^2\sinh^2\rho\,d\varphi^2
\right],
\label{eq:CD-conformal-metric}
\end{equation}
where
\begin{equation}
\Omega(\eta,\rho)
=
\frac{1}{
\cosh\rho+\cosh(2\eta/\alpha)
}
\label{eq:CD-conformal-factor-general}
\end{equation}
is the conformal factor. The region \(D\), equipped with this induced
metric, defines the Lorentzian diamond manifold \(\mathcal M_D\). The
spatial line element inside the brackets is that of the hyperbolic
plane \(H^2\) with curvature radius \(\alpha\). Thus, the line element
in the diamond chart is conformal to the product geometry
\(\mathbb R_\eta\times H^2\).

The metric inside the brackets in
Eq.~\eqref{eq:CD-conformal-metric} is independent of \(\eta\), so
translations in \(\eta\) are isometries of the product geometry.
Because the conformal factor depends on \(\eta\), the same
translations leave the Minkowski metric restricted to \(D\)
invariant only up to a local rescaling. Their generator is therefore
a conformal Killing vector
\cite{jacobson2019gravitational,arzano2020conformal,
eissa2025horizon}.
\begin{equation}
\xi
=
\partial_\eta
=
\left(
\frac{\alpha^2-t^2-r^2}{\alpha^2}
\right)\partial_t
-
\frac{2tr}{\alpha^2}\partial_r.
\label{eq:diamond-CKV-section-II}
\end{equation}
With this normalization, \(\xi\) coincides with \(\partial_t\) at the
center of the diamond. The constant-\(\rho\) curves shown in
Fig.~\ref{fig:CD-radial}(a) are therefore the integral curves of the
conformal flow generated by \(\xi\).

The causal character of this flow follows from its norm,
\begin{equation}
\xi\!\cdot\!\xi
=
-
\frac{
\left[\alpha^2-(t+r)^2\right]
\left[\alpha^2-(t-r)^2\right]
}{
\alpha^4
}.
\label{eq:diamond-CKV-norm-section-II}
\end{equation}
Inside \(D\), both factors in the numerator are positive, and
\(\xi\) is timelike. On the regular portions of the causal
boundary, one of the factors vanishes and the vector becomes null.
At the waist \(\Sigma_0\) and at the tips \(i_D^\pm\), the vector
itself vanishes. Therefore, the regular portions of the future and
past boundaries are conformal Killing horizons generated by
\(\xi\).

To describe the horizon limits of the chart explicitly, we introduce
the Minkowski null coordinates
\begin{equation}
U=t-r,
\qquad
V=t+r.
\label{eq:CD-null-coordinates}
\end{equation}
The future and past horizons are represented by \(V=\alpha\) and
\(U=-\alpha\), respectively. In terms of the diamond
coordinates, these null variables take the form
\begin{equation}
\begin{aligned}
U&=\alpha\tanh\!\left[\tfrac12\left(\tfrac{2\eta}{\alpha}-\rho\right)\right],\\
V&=\alpha\tanh\!\left[\tfrac12\left(\tfrac{2\eta}{\alpha}+\rho\right)\right].
\end{aligned}
\label{eq:UV-eta-rho}
\end{equation}

A generic point on \(\mathcal H^+\) is approached by keeping
\(2\eta/\alpha-\rho\) finite while taking the limit
\(2\eta/\alpha+\rho\to+\infty\), so that \(V\to\alpha\).
Similarly, a generic point on \(\mathcal H^-\) is approached by
keeping \(2\eta/\alpha+\rho\) finite while taking
\(2\eta/\alpha-\rho\to-\infty\), so that \(U\to-\alpha\).
Therefore, although the limit \(\rho\to\infty\) at fixed finite
\(\eta\) approaches the waist, generic points on the complete causal
boundary require correlated asymptotic limits of the two
coordinates.

In the null coordinates, the conformal Killing vector becomes
\begin{equation}
\xi
=
\frac{\alpha^2-U^2}{\alpha^2}\,\partial_U
+
\frac{\alpha^2-V^2}{\alpha^2}\,\partial_V.
\label{eq:CKV-null-form-section-II}
\end{equation}
For the conformal Killing horizon, the associated surface gravity
\(\kappa\) characterizes the scale of the horizon-generating flow and,
as will be established in the subsequent analysis, fixes the
thermodynamic scale of the diamond. It is defined on the
horizon through the proportionality between the gradient of
\(\xi^2\equiv
\xi\!\cdot\!\xi\)
and \(\xi_\mu\)
\cite{jacobson1993conformal}.
Evaluating this relation on the two branches gives
\begin{equation}
\nabla_\mu(\xi^2)
=
2\hat\epsilon\,\kappa\,\xi_\mu,
\qquad
\kappa=\frac{2}{\alpha},
\qquad
\hat\epsilon=\pm1,
\label{eq:surface-gravity-gradient-definition}
\end{equation}
where
\(\hat\epsilon=-1\) on \(\mathcal H^+\) and
\(\hat\epsilon=+1\) on \(\mathcal H^-\). The sign
\(\hat\epsilon\) records the branch dependence of this relation,
whereas the magnitude of the surface gravity is the same on both
horizons. For the normalization
\(\xi=\partial_\eta\), the diamond surface gravity
is therefore \(\kappa=2/\alpha\).

The diamond chart has therefore identified the conformal flow,
its horizon limits, and the corresponding surface gravity. It has
also shown that a surface of large constant \(\rho\) approaches the
Lorentzian causal boundary nonuniformly. We now analyze the geometric
region relevant for evaluating the gravitational boundary term in
Eq.~\eqref{eq:Euclidean-gravity-action}.
\subsection{Near-horizon geometry around the waist and \texorpdfstring{diamond temperature}{diamond temperature}}
\label{subsec:near-waist-temperature}

Since the diamond chart covers only the interior of \(D\), we
implement the stretched-horizon prescription by introducing the
timelike regulating hypersurface \(\rho=\rho_\circ\), with
\(\rho_\circ\gg1\), inside the diamond
\cite{PriceThorne1986,ParikhWilczek1998}. For finite values of
\(\eta\), the limit \(\rho\to\infty\) approaches the waist
\(\Sigma_0\). Therefore, the portion of the regulating surface
relevant for the boundary-term calculation is its near-waist region.

To describe the geometry in the vicinity of the waist, we introduce
the inward radial coordinate
\(X\equiv\alpha-r\). In the radial section, the waist is then
represented by \(t=X=0\). We also define
\begin{equation}
\ell
=
2\alpha e^{-\rho},
\qquad
\ell_\circ
=
2\alpha e^{-\rho_\circ},
\label{eq:near-waist-proper-coordinate}
\end{equation}
so that \(\ell\to0\) as \(\rho\to\infty\), while the regulating
hypersurface \(\rho=\rho_\circ\) is equivalently described by
\(\ell=\ell_\circ\). Expanding the coordinate map
\eqref{eq:CD-conformal-map-t}--\eqref{eq:CD-conformal-map-r} at
large \(\rho\), and using \(\kappa=2/\alpha\), gives at leading
order
\begin{equation}
t
\simeq
\ell\sinh(\kappa\eta),
\qquad
X
\simeq
\ell\cosh(\kappa\eta).
\label{eq:normal-rindler-coordinates}
\end{equation}

The omitted terms remain small in the near-waist patch,
\(X/\alpha\ll1\), which, through
Eq.~\eqref{eq:normal-rindler-coordinates}, becomes
\begin{equation}
\frac{\ell}{\alpha}\cosh(\kappa\eta)\ll1.
\label{eq:near-waist-validity-ell}
\end{equation}

\begin{figure*}[t]
\centering
\begin{minipage}{0.44\textwidth}
\centering
\includegraphics[width=0.8\linewidth]{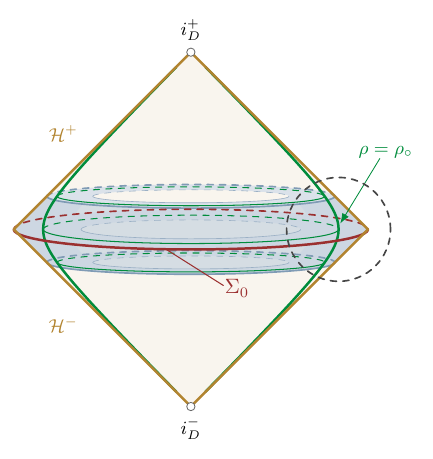}\par
\smallskip
(a)
\end{minipage}
\begin{minipage}{0.44\textwidth}
\centering
\includegraphics[width=\linewidth]{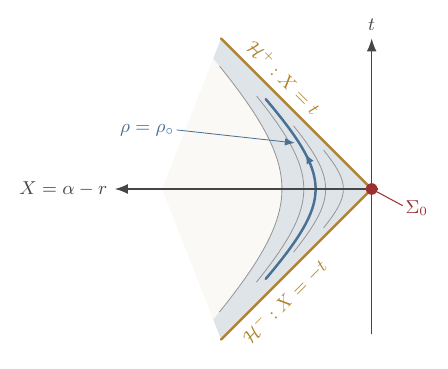}\par
\smallskip
(b)
\end{minipage}
\caption{
Analytic schematic of the regulating hypersurface and near-horizon geometry around the waist.
(a) The fixed-\(\rho\) regulator inside the
\((2+1)\)-dimensional causal diamond; the shaded band identifies the
near-waist region, and the dashed black circumference marks the local patch shown
in (b).
(b) The corresponding \(t\)-\(X\) section, with the regulator at
\(\rho=\rho_\circ\) and the local horizons at \(X=\pm t\).
}
\label{fig:CD-near-waist}
\end{figure*}

At each point of the waist, the coordinates \((t,X)\) span the
two-dimensional normal plane, while \(\varphi\) parametrizes the
direction tangent to \(\Sigma_0\). At the order retained in
Eq.~\eqref{eq:normal-rindler-coordinates}, curves of constant
\(\ell\) satisfy
\(X^2-t^2=\ell^2\). The near-waist portions of the future and past
horizons are correspondingly represented by the null lines
\(X=t\) and \(X=-t\), while the regulator
\(\rho=\rho_\circ\) is represented by the hyperbola
\(\ell=\ell_\circ\), as shown in
Fig.~\ref{fig:CD-near-waist}.

In terms of the inward coordinate \(X\), the Minkowski spacetime line element
takes the form
\(ds^2=-dt^2+dX^2+(\alpha-X)^2d\varphi^2\). Within the regime
\eqref{eq:near-waist-validity-ell}, the angular factor
\((\alpha-X)^2\) approaches \(\alpha^2\). Substituting
Eq.~\eqref{eq:normal-rindler-coordinates} then gives
\begin{equation}
ds^2_{(\mathrm{NH})}
\simeq
-\kappa^2\ell^2d\eta^2
+
d\ell^2
+
\alpha^2d\varphi^2.
\label{eq:CD-lorentzian-rindler-near-waist}
\end{equation}
Here and below, the label \((\mathrm{NH})\) denotes the leading
near-horizon approximation about the waist \(\Sigma_0\). 

The spacetime geometry normal to \(\Sigma_0\) is therefore locally
Rindler. Moreover, the unit coefficient of \(d\ell^2\) shows that
\(\ell\) is the proper normal distance from the waist at the order
retained. At the same order, the regulating hypersurface
\(\rho=\rho_\circ\) is therefore represented locally by the surface
of constant proper distance \(\ell=\ell_\circ\).

We now continue the Lorentzian diamond manifold \(\mathcal M_D\) to the
Euclidean diamond manifold \(\mathcal M_D^E\). Near the waist, this
continuation is implemented by setting \(\eta=-i\eta_E\) and
\(t_E\equiv it\). Equation
\eqref{eq:normal-rindler-coordinates} becomes, at the same order,

\begin{equation}
t_E
\simeq
\ell\sin(\kappa\eta_E),
\qquad
X
\simeq
\ell\cos(\kappa\eta_E).
\label{eq:normal-euclidean-polar-coordinates}
\end{equation}
Thus, \(\ell\) is the radial coordinate in the Euclidean normal
plane, while \(\kappa\eta_E\) is its angular coordinate. The
Euclidean continuation of
Eq.~\eqref{eq:CD-lorentzian-rindler-near-waist} is
\begin{equation}
ds_E^2
\simeq
d\ell^2
+
\kappa^2\ell^2d\eta_E^2
+
\alpha^2d\varphi^2.
\label{eq:CD-euclidean-Rindler}
\end{equation}

At \(\ell=0\), the circle generated by this normal angle collapses,
while \(\varphi\) continues to parametrize the waist \(\Sigma_0\).
Smoothness at this locus requires \(\kappa\eta_E\) to have period
\(2\pi\). Consequently, \(\eta_E\) must be
periodically identified with the regular Euclidean period
\(\beta_D\):
\begin{equation}
\eta_E
\sim
\eta_E+\beta_D,
\qquad
\beta_D
=
\frac{2\pi}{\kappa}
=
\pi\alpha.
\label{eq:CD-euclidean-period}
\end{equation}
The corresponding temperature of the conformal flow generated by
\(\xi=\partial_\eta\) is therefore
\begin{equation}
T_D
=
\frac{1}{\beta_D}
=
\frac{\kappa}{2\pi}
=
\frac{1}{\pi\alpha}.
\label{eq:CD-temperature}
\end{equation}

Having fixed the regular
Euclidean period and the corresponding diamond temperature,
we now evaluate the gravitational action carried by this regulating
hypersurface.

\subsection{Euclidean horizon action and equilibrium thermodynamics}
\label{subsec:euclidean-horizon-action}

After Euclidean continuation, the timelike regulator
\(\rho=\rho_\circ\) becomes the regulating tube
\(\ell=\ell_\circ\) surrounding the waist. At finite
\(\ell_\circ\), this tube is the surface on which the boundary term is
evaluated and bounds a tubular neighborhood with local product
structure
\begin{equation}
D^2_{\ell_\circ}\times\Sigma_0,
\label{eq:local-tubular-neighborhood}
\end{equation}
where \(D^2_{\ell_\circ}\) is the disk of radius \(\ell_\circ\) in
the Euclidean normal plane. As \(\rho_\circ\to\infty\), equivalently
\(\ell_\circ\to0\), the entire regulating tube collapses uniformly
onto the waist \(\Sigma_0\), reaching the Euclidean boundary limit.
Restricting the Euclidean line element
\eqref{eq:CD-euclidean-Rindler} to \(\ell=\ell_\circ\) gives
\begin{equation}
\left.ds_E^2\right|_{\ell=\ell_\circ}
=
\kappa^2\ell_\circ^2d\eta_E^2
+
\alpha^2d\varphi^2.
\label{eq:tubular-induced-line-element}
\end{equation}

At the order retained, the corresponding measure factor is
\(\sqrt h=\kappa\alpha\ell_\circ
[1+O(\ell_\circ/\alpha)]\). With the outward normal oriented toward
increasing \(\ell\), one has locally
\(n^\mu\partial_\mu=\partial_\ell\). Using
\(K=\nabla_\mu n^\mu\), the trace of the extrinsic curvature is
\(K=(1/\ell_\circ)[1+O(\ell_\circ/\alpha)]\). Consequently,
\begin{equation}
\sqrt h\,K
=
\kappa\alpha
\left[
1+O(\ell_\circ/\alpha)
\right].
\label{eq:tubular-boundary-product}
\end{equation}

Thus, the leading regulator dependence cancels in the product
\(\sqrt h\,K\). Since the regular bulk contribution of
\(\mathcal M_D^E\) vanishes, the complete on-shell Euclidean action is
obtained by evaluating the Gibbons-Hawking-York term on the
regulating tube and then taking the boundary limit
\(\ell_\circ\to0\):
\begin{align}
I_D^{E}
&\equiv
I_{\rm GHY}^{E}
\nonumber\\
&=
-\lim_{\ell_\circ\to0}
\frac{1}{8\pi G}
\int_0^{\beta_D}d\eta_E
\int_0^{2\pi}d\varphi\,
\sqrt h\,K
\nonumber\\
&=
-\frac{\beta_D\kappa}{8\pi G}A_0
=
-\frac{A_0}{4G}.
\label{eq:CD-smooth-horizon-action}
\end{align}

At the regular period, the standard semiclassical relation between
the Euclidean action and the canonical free energy
\cite{GibbonsHawking1977,hawking1979path}, introduced in
Eq.~\eqref{eq:gravitational-path-integral}, becomes
\begin{equation}
\begin{aligned}
Z_{\rm grav}(\beta_D)&\simeq e^{-I_D^E}=e^{-\beta_D F_D},\\
I_D^E&=\beta_D F_D=\beta_D E_D-S_D.
\end{aligned}
\label{eq:CD-saddle-thermodynamic-relation}
\end{equation}
Here, \(F_D\), \(E_D\), and
\(S_D\) denote the free energy, equilibrium
geometric energy, and entropy associated with the regulated diamond
saddle. Following Laflamme's Euclidean Rindler
analysis~\cite{Laflamme1987}, the absence of a mass term associated
with the diamond geometry implies that its equilibrium geometric
energy vanishes, \(E_D=0\).
Equation~\eqref{eq:CD-saddle-thermodynamic-relation} therefore reduces to
\(I_D^{E}=-S_D\).

The same purely entropic identification admits a complementary local
interpretation through the puncture prescription of Banks, Draper,
and Farkas~\cite{banks2021path}. On the regulating surface
\(\ell=\ell_\circ\), the proper circumference of the regular
Euclidean-time orbit is
\begin{equation}
\begin{aligned}
\beta_{\rm proper}(\ell_\circ)
&\equiv
\int_0^{\beta_D}d\eta_E\,
\sqrt{g_{\eta_E\eta_E}}
\\
&\simeq
\kappa\ell_\circ\beta_D
=
2\pi\ell_\circ
\longrightarrow
0.
\end{aligned}
\label{eq:CD-proper-period-puncture}
\end{equation}
The local temperature therefore diverges as the regulating surface
approaches the puncture. In this limit, the partition function counts
the gravitational states associated with the diamond, so the
thermodynamic relation reduces again to
\(I_D^{E}=-S_D\)~\cite{banks2021path}. Here,
\(\beta_{\rm proper}\) is the proper length of the same Euclidean
orbit whose coordinate period is \(\beta_D=2\pi/\kappa\).

Collecting these results, the equilibrium data of the unperturbed
diamond horizon are
\begin{equation}
T_D
=
\frac{1}{\pi\alpha},
\qquad
S_D
=
\frac{A_0}{4G},
\qquad
E_D
=
0.
\label{eq:CD-energy-entropy}
\end{equation}
\section{Mechanical laws of \texorpdfstring{diamond horizons}{diamond horizons}}
\label{sec:mechanical-laws}

The preceding Euclidean analysis established the equilibrium
temperature and entropy of the unperturbed diamond through the
regularity and boundary structure of the geometry surrounding the
waist. We now perturb the corresponding Lorentzian near-waist region
and determine its mechanical response.

The mechanical analysis below is also
carried out explicitly in \((2+1)\) spacetime dimensions. Its
extension to spherical diamonds in general \(d\geq3\)
spacetime dimensions is presented in
Appendix.

\subsection{Near-waist geometry and Killing-energy flux}
\label{subsec:perturbative-horizon-setup}

To formulate the mechanical process, we first isolate the portion of
the future horizon through which the stress-energy perturbation will
pass. The null coordinates \(U\) and \(V\), defined in
Eq.~\eqref{eq:CD-null-coordinates}, represent the future horizon by
\(V=\alpha\), as stated in
Eq.~\eqref{eq:CD-null-boundaries-plus}. At the waist, where \(t=0\)
and \(r=\alpha\), the same definitions give \(U=-\alpha\) and
\(V=\alpha\). We select an angular interval
\(\varphi_1\leq\varphi\leq\varphi_2\) on the waist and denote the
corresponding initial arc by
\begin{equation}
C_{\rm i}
=
\left\{
V=\alpha,\;
U=-\alpha,\;
\varphi_1\leq\varphi\leq\varphi_2
\right\}
\subset\Sigma_0.
\label{eq:mechanical-initial-arc}
\end{equation}
Each value of \(\varphi\) identifies one radial null generator issuing
from this arc. Following these generators toward the future tip
produces the horizon bundle on which the mechanical response will be
studied.

An affine parameter along the generators is needed both to locate the
stress-energy flux and to measure the resulting change in their
separation. Since \(U\) varies linearly along the radial null
geodesics of \(\mathcal H^+\), we introduce the normalized parameter
\begin{equation}
\lambda
\equiv
\left.
\frac{U+\alpha}{2}
\right|_{\mathcal H^+}.
\label{eq:mechanical-affine-parameter}
\end{equation}
This parameter vanishes on \(C_{\rm i}\), since \(U=-\alpha\) at the
waist. On the future horizon, \(V=\alpha\) and
\(U=2\lambda-\alpha\), so the reference generators are described by
\begin{equation}
t(\lambda)=\lambda,
\qquad
r(\lambda)=\alpha-\lambda,
\qquad
0\leq\lambda<\alpha.
\label{eq:mechanical-reference-generator-embedding}
\end{equation}
Thus, increasing \(\lambda\) moves each generator forward in time and
radially inward from the waist toward the future tip. We denote its
tangent vector by \(k\). Since \(\varphi\) remains constant along each
radial generator, the only nonzero coordinate components of \(k\) are
those in the \((t,r)\) plane:
\begin{equation}
\left(k^t,k^r\right)
\equiv
\left(
\frac{dt}{d\lambda},
\frac{dr}{d\lambda}
\right)
=
(1,-1).
\label{eq:mechanical-affine-generator-components}
\end{equation}
In the coordinate basis
\(\{\partial_t,\partial_r\}\), the tangent vector itself is therefore
\begin{equation}
\begin{aligned}
k&\equiv k^\mu\partial_\mu=\frac{d}{d\lambda}
=2\partial_U=\partial_t-\partial_r,\\
k^\mu k_\mu&=0,\qquad k^\nu\nabla_\nu k^\mu=0.
\end{aligned}
\label{eq:mechanical-affine-generator}
\end{equation}
The positive \(t\) component and negative \(r\) component show that
\(k\) is future directed and radially inward pointing. The condition
\(k^\mu k_\mu=0\) shows that it is null, while
\(k^\nu\nabla_\nu k^\mu=0\) verifies that it is geodesic and that
\(\lambda\) is an affine parameter.

\begin{figure}[t]
\centering
\includegraphics[width=0.91\columnwidth]{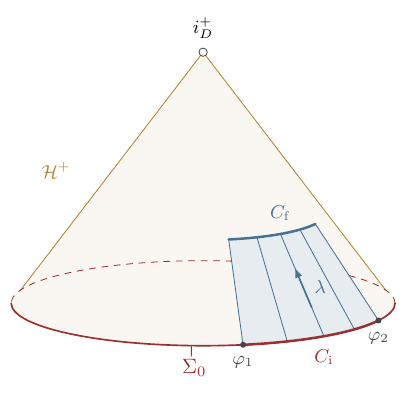}
\caption{
Schematic null-generator bundle on \(\mathcal H^+\) between the waist arc
\(C_{\rm i}\) and the final cut \(C_{\rm f}\).
}
\label{fig:mechanical-null-patch}
\end{figure}

We follow the reference bundle from \(C_{\rm i}\) to the arc
\begin{equation}
C_{\rm f}
=
\left\{
V=\alpha,\;
\lambda=\lambda_f,\;
\varphi_1\leq\varphi\leq\varphi_2
\right\}
\subset\mathcal H^+.
\label{eq:mechanical-final-arc}
\end{equation}
At fixed \(\lambda\), the cut lies at radius
\(r=\alpha-\lambda\). Its exact reference length element is therefore
\begin{equation}
dL_{\rm ref}(\lambda)
=
(\alpha-\lambda)\,d\varphi.
\label{eq:mechanical-reference-transverse-measure}
\end{equation}
The decrease of this element with \(\lambda\) describes the
geometrical contraction of the future-horizon cuts as their
generators approach the tip, as illustrated in
Fig.~\ref{fig:mechanical-null-patch}.

We choose the final cut within the near-waist regime,
\begin{equation}
\frac{\lambda_f}{\alpha}\ll1.
\label{eq:mechanical-near-waist-interval}
\end{equation}
Since \(0\leq\lambda\leq\lambda_f\), the entire selected bundle
remains within that region. Along the reference generators,
\(X\equiv\alpha-r=\lambda\), so
Eq.~\eqref{eq:mechanical-near-waist-interval} is precisely the
horizon restriction of the near-waist condition \(X/\alpha\ll1\)
introduced in Sec.~\ref{subsec:near-waist-temperature}. Expanding the exact
diamond-time generator at the corresponding order gives
\begin{equation}
\xi
\simeq
\xi^{(\mathrm{NH})}
=
\kappa\left(X\partial_t+t\partial_X\right).
\label{eq:mechanical-local-boost-generator}
\end{equation}
This is the boost Killing vector of the local Rindler geometry. Since
\(t=X=\lambda\) and \(k=\partial_t+\partial_X\) on
\(\mathcal H^+\),
\begin{equation}
\left.\xi^{(\mathrm{NH})\,\mu}\right|_{\mathcal H^+}
=
\kappa\lambda k^\mu.
\label{eq:mechanical-local-boost-affine-relation}
\end{equation}
At the same order, the reference transverse measure reduces to its
waist value,
\(dL_{\rm ref}(\lambda)\simeq dL_{\rm ref}(0)
=\alpha\,d\varphi\).

We consider a weak stress-energy perturbation about the reference
diamond and write
\(T_{\mu\nu}^{\rm pert}
=T_{\mu\nu}^{\rm ref}+\delta T_{\mu\nu}\), with
\(T_{\mu\nu}^{\rm ref}=0\), working to linear order in the
perturbation. Because the reference stress tensor vanishes, the
horizon-crossing flux at this order can be evaluated using the
generators and surface element of the reference bundle. In a semiclassical treatment,
\(\delta T_{\mu\nu}\equiv
\delta\langle\hat T_{\mu\nu}\rangle_{\rm ren}\). Renormalization is
required because the expectation value of the local composite
operator \(\hat T_{\mu\nu}\) is ultraviolet divergent before the
standard subtractions are performed
\cite{BirrellDavies1982,Wald1994}.

The local boost symmetry defines the Killing current
\begin{equation}
\delta J^\mu
\equiv
\delta T^\mu{}_{\nu}\,\xi_{(\mathrm{NH})}^{\nu}.
\label{eq:mechanical-diamond-energy-current}
\end{equation}
The null-energy projection of the stress-energy perturbation along
the horizon generators is
\(\delta T_{kk}\equiv\delta T_{\mu\nu}k^\mu k^\nu\). For the
classical matter perturbations considered in the positivity analysis
below, we assume the null energy condition. Along the selected horizon
generators, this gives
\begin{equation}
\delta T_{kk}(\lambda,\varphi)\geq0.
\label{eq:mechanical-null-energy-condition}
\end{equation}
The selected null strip is parametrized by
\((\lambda,\varphi)\). We choose its orientation so that
\(\delta T_{kk}>0\) contributes positively to the flux. Its directed
surface element at the retained order is
\begin{equation}
d\Sigma_\mu
=
k_\mu\,d\lambda\,dL_{\rm ref}(0).
\label{eq:mechanical-null-strip-surface-element}
\end{equation}
The Killing-energy flux is therefore
\begin{align}
\delta E
&\equiv
\int
\delta J^\mu\,d\Sigma_\mu
\nonumber
\\
&=
\kappa
\int_{C_{\rm i}}dL_{\rm ref}(0)
\int_0^{\lambda_f}d\lambda\,
\lambda\,\delta T_{kk}
\nonumber
\\
&=
\kappa\alpha
\int_{\varphi_1}^{\varphi_2}d\varphi
\int_0^{\lambda_f}d\lambda\,
\lambda\,\delta T_{kk}.
\label{eq:mechanical-Killing-energy-flux}
\end{align}
The Lorentzian flux \(\delta E\) should not be confused with the
equilibrium geometric energy \(E_D=0\) introduced in
Sec.~\ref{sec:equilibrium-thermodynamics}. The affine limits delimit
the selected bundle, while
\(\delta T_{kk}(\lambda,\varphi)\) determines the distribution of
matter crossing it. The factor \(\lambda\) reflects the linear
growth of the boost generator away from the waist.

\subsection{Raychaudhuri response and area variation at a common affine endpoint}
\label{subsec:raychaudhuri-area-response}

The preceding subsection defined the Killing-energy flux
\(\delta E\) through the selected near-waist bundle. The same
null-energy projection \(\delta T_{kk}\) also sources the gravitational
response
through the linearized Einstein equation and changes the transverse
separation of neighboring generators. We now determine this geometric response by comparing the reference
horizon bundle with the corresponding null bundle in the perturbed
geometry. The two bundles are matched on the initial cut
\(C_{\rm i}\).

\begin{figure}[t]
\centering
\includegraphics[width=0.98\columnwidth]{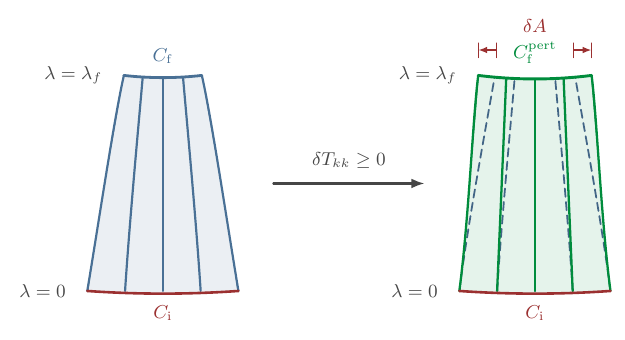}
\caption{
Schematic comparison of the reference horizon bundle and the
corresponding perturbed null bundle matched at \(C_{\rm i}\), with
final cuts \(C_{\rm f}\) and \(C_{\rm f}^{\rm pert}\) defining the
area variation \(\delta A\) at \(\lambda=\lambda_f\). Dashed
curves in the perturbed panel reproduce the reference generators,
whereas solid curves show the perturbed bundle.
}
\label{fig:mechanical-reference-perturbed}
\end{figure}

At the common affine endpoint
\(\lambda=\lambda_f\), we denote the reference final cut by
\(C_{\rm f}\) and its perturbed counterpart by
\(C_{\rm f}^{\rm pert}\); their area difference defines the
variation \(\delta A\) at that common affine endpoint. The null
Raychaudhuri equation relates
\(\delta T_{kk}\) to the change in the expansion of the generators,
from which their change in transverse separation can be obtained
\cite{Raychaudhuri1955,Wald1984}, as summarized in
Fig.~\ref{fig:mechanical-reference-perturbed}.

The parameters \((\lambda,\varphi)\) identify corresponding
generators in the two bundles. On each perturbed generator,
\(\lambda\) is chosen to be affine, to vanish at \(C_{\rm i}\), and
to have the same normalization there as the reference parameter. We also assume that neighboring generators do not cross and that no
generator enters or leaves the selected bundle between the initial
and corresponding final cuts
\cite{AmselMarolfVirmani2008}.

At fixed \(\lambda\), each bundle is cut by a one-dimensional arc.
Its transverse geometry is therefore completely described by the
local length separating neighboring generators. We denote the length
per unit angular label by \(\mathcal L_{\rm ref}(\lambda)\) for the
reference bundle and by
\(\mathcal L_{\rm pert}(\lambda,\varphi)\) for the perturbed bundle:
\begin{equation}
dL_{\rm ref}
=
\mathcal L_{\rm ref}(\lambda)\,d\varphi,
\qquad
dL_{\rm pert}
=
\mathcal L_{\rm pert}(\lambda,\varphi)\,d\varphi.
\label{eq:mechanical-transverse-line-elements}
\end{equation}
The expansion measures the fractional rate of change of this local
length along a generator,
\begin{equation}
\theta
\equiv
\frac{1}{\mathcal L}
\frac{d\mathcal L}{d\lambda},
\label{eq:mechanical-expansion-definition}
\end{equation}
where the derivative is taken at fixed \(\varphi\). Positive expansion
means that neighboring generators move apart, whereas negative
expansion means that they converge.

Although the selected interval lies near the waist, the cuts of the
exact reference horizon continue to contract as their generators
approach the future tip. We retain this background contraction until
its contribution has been separated from the perturbative response.
Equation~\eqref{eq:mechanical-reference-transverse-measure} gives
\begin{equation}
\mathcal L_{\rm ref}(\lambda)
=
\alpha-\lambda,
\qquad
\theta_{\rm ref}
=
-\frac{1}{\alpha-\lambda}.
\label{eq:mechanical-reference-expansion}
\end{equation}
The negative sign describes the narrowing of the reference cuts as
\(\lambda\) increases.

We denote the corresponding variation of the expansion by
\(\delta\theta\equiv\theta_{\rm pert}-\theta_{\rm ref}\). In \((2+1)\) dimensions, the transverse cut is one dimensional, so
its deformation has no traceless part and the shear vanishes
identically. The twist also vanishes because the generators form a
null hypersurface. The null Raychaudhuri equation therefore reduces
to
\begin{equation}
\frac{d\theta}{d\lambda}
=
-\theta^2
-
R_{\mu\nu}k^\mu k^\nu.
\label{eq:mechanical-raychaudhuri-2p1}
\end{equation}
For the flat reference geometry,
\(d\theta_{\rm ref}/d\lambda=-\theta_{\rm ref}^2\). Writing the
curvature of the perturbed geometry as
\(R_{\mu\nu}^{\rm pert}
=R_{\mu\nu}^{\rm ref}+\delta R_{\mu\nu}\), and applying
Eq.~\eqref{eq:mechanical-raychaudhuri-2p1} to the perturbed
congruence, linearization about the reference congruence gives
\begin{equation}
\frac{d\,\delta\theta}{d\lambda}
=
-2\theta_{\rm ref}\,\delta\theta
-
\delta R_{\mu\nu}k^\mu k^\nu.
\label{eq:mechanical-linearized-raychaudhuri-curvature}
\end{equation}

Contracting the linearized Einstein equation twice with the reference
null tangent gives
\begin{equation}
\begin{aligned}
\left(\delta R_{\mu\nu}
-\tfrac12 g^{\rm ref}_{\mu\nu}\delta R\right)k^\mu k^\nu
&=\delta R_{\mu\nu}k^\mu k^\nu\\
&=8\pi G\,\delta T_{\mu\nu}k^\mu k^\nu\\
&=8\pi G\,\delta T_{kk}.
\end{aligned}
\label{eq:mechanical-null-einstein-equation}
\end{equation}
Here, \(g_{\mu\nu}^{\rm ref}\) is the reference Minkowski metric,
while \(\delta R\) denotes the corresponding Ricci-scalar variation.
The term proportional to \(\delta R\) vanishes because
\(g_{\mu\nu}^{\rm ref}k^\mu k^\nu=0\). Since the reference geometry
is flat and its stress-energy tensor vanishes,
\(R_{\mu\nu}^{\rm ref}=T_{\mu\nu}^{\rm ref}=0\), variations of the
tangent do not contribute to the linearized contraction.

Using Eqs.~\eqref{eq:mechanical-reference-expansion} and
\eqref{eq:mechanical-null-einstein-equation}, the linearized
Raychaudhuri equation becomes
\begin{equation}
\frac{d\,\delta\theta}{d\lambda}
=
\frac{2}{\alpha-\lambda}\,\delta\theta
-
8\pi G\,\delta T_{kk}.
\label{eq:mechanical-linearized-raychaudhuri-exact}
\end{equation}

The final cut is also taken after the matter flux through the selected
bundle has ceased, while still satisfying
\(\lambda_f/\alpha\ll1\). We then impose the final matching condition
\begin{equation}
\delta\theta(\lambda_f,\varphi)=0,
\label{eq:mechanical-final-expansion-matching}
\end{equation}
which requires the perturbed and reference bundles to have the same
instantaneous contraction rate on their respective final cuts. This
condition selects the solution of the linearized Raychaudhuri equation.

Equation~\eqref{eq:mechanical-linearized-raychaudhuri-exact} can be
written as
\begin{equation}
\frac{d}{d\lambda}
\left[
(\alpha-\lambda)^2\delta\theta
\right]
=
-8\pi G
(\alpha-\lambda)^2
\delta T_{kk}.
\label{eq:mechanical-raychaudhuri-integrating-factor}
\end{equation}
Integrating this equation from \(\lambda\) to \(\lambda_f\) gives
\begin{equation}
\begin{aligned}
&(\alpha-\lambda_f)^2\delta\theta(\lambda_f,\varphi)
-(\alpha-\lambda)^2\delta\theta(\lambda,\varphi)\\
&\qquad=-8\pi G\int_\lambda^{\lambda_f}d\lambda'\,
(\alpha-\lambda')^2\delta T_{kk}(\lambda',\varphi).
\end{aligned}
\label{eq:mechanical-raychaudhuri-integrated}
\end{equation}
Using the final matching condition
\(\delta\theta(\lambda_f,\varphi)=0\), we obtain
\begin{equation}
\delta\theta(\lambda,\varphi)
=
\frac{8\pi G}{(\alpha-\lambda)^2}
\int_\lambda^{\lambda_f}
d\lambda'\,
(\alpha-\lambda')^2
\delta T_{kk}(\lambda',\varphi).
\label{eq:mechanical-expansion-response-exact}
\end{equation}
Under the null energy condition in
Eq.~\eqref{eq:mechanical-null-energy-condition}, every factor in
Eq.~\eqref{eq:mechanical-expansion-response-exact} is nonnegative.
Therefore,
\(\delta\theta(\lambda,\varphi)\geq0\) on cuts preceding
\(C_{\rm f}\). This does not mean that the perturbed expansion is
itself positive. Rather, the perturbed bundle contracts less rapidly
than the reference bundle. The focusing effect is seen directly from
Eq.~\eqref{eq:mechanical-raychaudhuri-integrating-factor}: under the
same condition, the quantity
\((\alpha-\lambda)^2\delta\theta\) decreases as \(\lambda\)
increases. With the final matching condition imposed, it vanishes at
\(\lambda=\lambda_f\), where the perturbed and reference expansions
agree on their corresponding final cuts.

The quantity \(\delta\theta\) measures the perturbation of this
contraction rate, rather than the accumulated change in the actual
separation between neighboring generators. To determine that
separation, we write
\(\mathcal L_{\rm pert}
=\mathcal L_{\rm ref}+\delta\mathcal L\). The two bundles are matched
intrinsically on the initial arc, so that
\(\delta\mathcal L(0,\varphi)=0\). Any nonzero value of
\(\delta\mathcal L\) on a later cut is therefore generated during the
process.

Linearizing
Eq.~\eqref{eq:mechanical-expansion-definition} gives
\begin{equation}
\frac{d\,\delta\mathcal L}{d\lambda}
=
\theta_{\rm ref}\,\delta\mathcal L
+
\mathcal L_{\rm ref}\,\delta\theta.
\label{eq:mechanical-linearized-length-evolution}
\end{equation}
The first term describes the evolution of an already existing length
difference under the contraction of the reference horizon, while the
second is the additional change generated by the perturbation of the
expansion. Using Eq.~\eqref{eq:mechanical-reference-expansion}, this
equation becomes
\begin{equation}
\frac{d}{d\lambda}
\left(
\frac{\delta\mathcal L}{\alpha-\lambda}
\right)
=
\delta\theta.
\label{eq:mechanical-fractional-length-evolution}
\end{equation}
Integrating from the initial cut to the final cut gives
\begin{equation}
\frac{\delta\mathcal L(\lambda_f,\varphi)}
{\alpha-\lambda_f}
-
\frac{\delta\mathcal L(0,\varphi)}{\alpha}
=
\int_0^{\lambda_f}
d\lambda\,
\delta\theta(\lambda,\varphi).
\end{equation}
Since \(\delta\mathcal L(0,\varphi)=0\), the accumulated local length
variation on the final cut is
\begin{equation}
\delta\mathcal L(\lambda_f,\varphi)
=
(\alpha-\lambda_f)
\int_0^{\lambda_f}
d\lambda\,
\delta\theta(\lambda,\varphi).
\label{eq:mechanical-final-local-length-response}
\end{equation}

In \((2+1)\) dimensions, the codimension-two measure is a length. We
continue to denote it by \(A\) in order to maintain the notation used
in the equilibrium area law. The area variation at the common affine
endpoint is defined by comparing the perturbed and reference cuts:
\begin{align}
\delta A
&\equiv
A(C_{\rm f}^{\rm pert})
-
A(C_{\rm f})
\nonumber
\\
&=
\int_{\varphi_1}^{\varphi_2}
d\varphi\,
\delta\mathcal L(\lambda_f,\varphi).
\label{eq:mechanical-endpoint-area-definition}
\end{align}
Because both terms are evaluated on corresponding cuts at the same
affine endpoint \(\lambda_f\), this difference removes the ordinary
contraction undergone by the reference horizon between
\(C_{\rm i}\) and \(C_{\rm f}\), leaving only the response produced
by the perturbation.

Substituting
Eq.~\eqref{eq:mechanical-expansion-response-exact} into
Eq.~\eqref{eq:mechanical-final-local-length-response} gives
\begin{equation}
\begin{aligned}
\delta\mathcal L(\lambda_f,\varphi)
&=8\pi G(\alpha-\lambda_f)
\int_0^{\lambda_f}\frac{d\lambda}{(\alpha-\lambda)^2}\\
&\quad\times\int_\lambda^{\lambda_f}d\lambda'\,
(\alpha-\lambda')^2\delta T_{kk}(\lambda',\varphi).
\end{aligned}
\label{eq:mechanical-local-length-double-integral}
\end{equation}
The integration region is
\(0\leq\lambda\leq\lambda'\leq\lambda_f\). Interchanging the order of
integration therefore gives
\begin{equation}
\begin{aligned}
\delta\mathcal L(\lambda_f,\varphi)
&=8\pi G(\alpha-\lambda_f)\int_0^{\lambda_f}d\lambda'\,
(\alpha-\lambda')^2\delta T_{kk}(\lambda',\varphi)\\
&\quad\times\int_0^{\lambda'}\frac{d\lambda}{(\alpha-\lambda)^2}.
\end{aligned}
\end{equation}
The inner integral is
\begin{equation}
\int_0^{\lambda'}
\frac{d\lambda}{(\alpha-\lambda)^2}
=
\frac{1}{\alpha-\lambda'}
-
\frac{1}{\alpha}
=
\frac{\lambda'}{\alpha(\alpha-\lambda')}.
\end{equation}
It follows that
\begin{equation}
\delta\mathcal L(\lambda_f,\varphi)
=
\frac{8\pi G(\alpha-\lambda_f)}{\alpha}
\int_0^{\lambda_f}
d\lambda\,
\lambda(\alpha-\lambda)
\delta T_{kk}(\lambda,\varphi).
\end{equation}
Integrating this local response over the angular interval then yields
\begin{equation}
\delta A
=
\frac{8\pi G(\alpha-\lambda_f)}{\alpha}
\int_{\varphi_1}^{\varphi_2}
d\varphi
\int_0^{\lambda_f}
d\lambda\,
\lambda(\alpha-\lambda)
\delta T_{kk}.
\label{eq:mechanical-endpoint-area-response-exact}
\end{equation}
The factors involving \(\alpha-\lambda\) retain the contraction of
the reference horizon throughout the selected interval.

For \(0\leq\lambda\leq\lambda_f\ll\alpha\), the geometric factor
satisfies
\(\frac{\alpha-\lambda_f}{\alpha}(\alpha-\lambda)
=
\alpha[1+O(\lambda_f/\alpha)]\)
uniformly across the selected bundle. Keeping the leading near-waist
term in Eq.~\eqref{eq:mechanical-endpoint-area-response-exact} gives
\begin{align}
\delta A
&\simeq
8\pi G\alpha
\int_{\varphi_1}^{\varphi_2}
d\varphi
\int_0^{\lambda_f}
d\lambda\,
\lambda\,\delta T_{kk}
\nonumber
\\
&=
8\pi G
\int_{C_{\rm i}}
dL_{\rm ref}(0)
\int_0^{\lambda_f}
d\lambda\,
\lambda\,\delta T_{kk}.
\label{eq:mechanical-endpoint-area-response-near-waist}
\end{align}
Thus, the same affine-weighted integral of \(\delta T_{kk}\) that
defines \(\delta E\) in
Eq.~\eqref{eq:mechanical-Killing-energy-flux} determines the leading
near-waist area response.

\subsection{Mechanical laws}
\label{subsec:mechanical-laws}

\subsubsection{First law: local energy--entropy balance}
\label{subsec:first-law-local-balance}

The Killing-energy flux \(\delta E\) and the area response
\(\delta A\) have been obtained from the same perturbation by two
independent routes. As shown in
Eqs.~\eqref{eq:mechanical-Killing-energy-flux} and
\eqref{eq:mechanical-endpoint-area-response-near-waist}, at the
leading order retained in the near-waist regime, both quantities
contain the same affine-weighted integral of \(\delta T_{kk}\).
Consequently,
\begin{equation}
\delta E
\simeq
\frac{\kappa}{8\pi G}\,\delta A.
\label{eq:mechanical-local-energy-area-relation}
\end{equation}

Equation~\eqref{eq:CD-energy-entropy} established the equilibrium
area law \(S_D=A_0/(4G)\) for the unperturbed diamond. In Einstein
gravity, the same coefficient \(1/(4G)\) supplies the local entropy
density associated with transverse horizon measure
\cite{Jacobson1995}. The area response of the selected horizon bundle
therefore corresponds to the entropy response
\begin{equation}
\delta S_{\rm hor}
\equiv
\frac{\delta A}{4G}.
\label{eq:mechanical-local-entropy-response}
\end{equation}
The scale \(\alpha\), and hence the equilibrium waist measure
\(A_0\), remains fixed throughout the mechanical process; \(\delta A\)
is instead the final-cut area response of the selected horizon bundle.

The temperature associated with the same normalized diamond-time
flow was fixed by Euclidean regularity in
Eq.~\eqref{eq:CD-temperature},
\(T_D=\beta_D^{-1}=\kappa/(2\pi)=1/(\pi\alpha)\).
Combining this result with
Eqs.~\eqref{eq:mechanical-local-energy-area-relation} and
\eqref{eq:mechanical-local-entropy-response} gives
\begin{equation}
\delta E
\simeq
T_D\,\delta S_{\rm hor}.
\label{eq:mechanical-local-first-law}
\end{equation}
This is the local physical-process first law for the selected horizon
bundle within the linearized near-waist regime.

\subsubsection{Second law: positivity of the local entropy response}
\label{subsec:second-law-local-increase}

The area response in
Eq.~\eqref{eq:mechanical-endpoint-area-response-exact} also determines
the sign of \(\delta A\). Throughout the selected interval,
\(0\leq\lambda\leq\lambda_f<\alpha\), all geometric factors in its
kernel are nonnegative. The null energy condition in
Eq.~\eqref{eq:mechanical-null-energy-condition} therefore implies
\begin{equation}
\delta A\geq0.
\label{eq:mechanical-local-area-increase}
\end{equation}
The inequality is strict when \(\delta T_{kk}>0\) on a region of
nonzero measure away from \(\lambda=0\).

Under the comparison at the common affine endpoint
\(\lambda=\lambda_f\),
\(\delta A=A(C_{\rm f}^{\rm pert})
-A(C_{\rm f})\). Thus, under this condition, the perturbed
final cut has length no smaller than that of the corresponding
reference cut.
The reference horizon continues to contract toward the future tip;
the inequality measures the additional response produced by the
perturbation relative to that background evolution.

The entropy assignment of
Eq.~\eqref{eq:mechanical-local-entropy-response} then gives
\begin{equation}
\delta S_{\rm hor}
=
\frac{\delta A}{4G}
\geq0.
\label{eq:mechanical-local-entropy-increase}
\end{equation}
This is the local second-law statement for the selected near-waist
process.

\subsubsection{Zeroth law and geometric third-law statement for finite \texorpdfstring{diamonds}{diamonds}}
\label{subsec:zeroth-law-zero-temperature-limit}

The preceding analysis concerned a local physical process on a
near-waist portion of \(\mathcal H^+\). The zeroth law instead
characterizes the equilibrium geometry of the complete unperturbed
diamond. We therefore leave the perturbed bundle and return to
the exact diamond geometry defined by
Eq.~\eqref{eq:causal-diamond-domain}, with horizons given by
Eqs.~\eqref{eq:CD-null-boundaries-plus} and
\eqref{eq:CD-null-boundaries-minus}.

The equilibrium flow is generated by
\(\xi=\partial_\eta\), whose normalization is fixed by the
diamond-time coordinate. As established in
Eq.~\eqref{eq:surface-gravity-gradient-definition}, this flow
satisfies
\begin{equation}
\nabla_\mu\!\left(\xi^2\right)
=
2\hat\epsilon\,\kappa\,\xi_\mu,
\qquad
\kappa=\frac{2}{\alpha},
\label{eq:mechanical-zeroth-law-definition}
\end{equation}
with \(\hat\epsilon=-1\) on \(\mathcal H^+\) and
\(\hat\epsilon=+1\) on \(\mathcal H^-\). The sign records the
opposite orientations of the two horizon branches, while the
magnitude \(\kappa\) is independent of the horizon coordinates.
Therefore, a diamond with fixed temporal extent
\(2\alpha\) possesses a single conformal surface gravity on the
regular portions of both horizons, establishing the zeroth law
directly in the present geometry.

Euclidean regularity assigns to the same normalized flow the period
and equilibrium temperature
\begin{equation}
\beta_D
=
\frac{2\pi}{\kappa}
=
\pi\alpha,
\qquad
T_D
=
\frac{1}{\beta_D}
=
\frac{\kappa}{2\pi}
=
\frac{1}{\pi\alpha}.
\label{eq:mechanical-zeroth-law-temperature}
\end{equation}
Thus, the uniform surface gravity determines a single global
temperature parameter associated with the diamond-time flow.

The dependence on \(\alpha\) also determines the zero-temperature
limit of the family. Every finite diamond has
\(0<\alpha<\infty\), and therefore \(\kappa>0\) and \(T_D>0\).
The temperature approaches zero only as the temporal extent
\(2\alpha\) diverges,
\begin{equation}
\begin{aligned}
\alpha\longrightarrow\infty&\quad\Longrightarrow\quad
\kappa\longrightarrow0,\\
T_D&\longrightarrow0,\qquad
\beta_D\longrightarrow\infty.
\end{aligned}
\label{eq:mechanical-zero-temperature-limit}
\end{equation}
This limit changes the causal region itself. Writing the family as
\(D_\alpha=\{(t,r,\varphi):|t|+r<\alpha\}\), every point with finite
\(t\) and \(r\) belongs to \(D_\alpha\) once
\(\alpha>|t|+r\). Hence,
\begin{equation}
\lim_{\alpha\to\infty}D_\alpha
=
\mathbb{R}^{2,1}.
\label{eq:mechanical-diamond-minkowski-limit}
\end{equation}
The zero-temperature limit therefore lies outside the family of
finite diamonds: as \(\alpha\to\infty\), the waist and the
tips recede to infinity and the finite causal boundary disappears,
leaving the complete Minkowski spacetime. This provides the
corresponding geometric third-law statement for the finite-diamond
family: zero
temperature is approached only in the limit in which the finite
causal horizon disappears.

\section{Conclusions}
\label{sec:conclusions}

The analysis developed in this article shows that the equilibrium
thermodynamics and local mechanical response of a causal-diamond
horizon arise from complementary aspects of the same finite causal
geometry.

First, Euclidean regularity of the normalized diamond-time flow fixes
\begin{equation}
T_D
=
\frac{\kappa}{2\pi}
=
\frac{1}{\pi\alpha},
\end{equation}
while the regulated horizon contribution to the Euclidean action,
together with the vanishing equilibrium geometric energy \(E_D=0\),
gives
\begin{equation}
S_D
=
\frac{A_0}{4G}.
\end{equation}
Thus, the equilibrium temperature and entropy of the causal diamond
are determined by the regularity and boundary structure of its
Euclidean geometry.

Second, the Lorentzian analysis determines the local mechanical
response of the horizon to matter. For a weak perturbation crossing a
selected near-waist portion of the future horizon, the Killing-energy
flux and the Raychaudhuri area response are governed, at leading
near-waist order, by the same affine-weighted stress-energy flux.
Their relation gives
\begin{equation}
\delta E
\simeq
\frac{\kappa}{8\pi G}\,\delta A
=
T_D\,\delta S_{\rm hor},
\qquad
\delta S_{\rm hor}
=
\frac{\delta A}{4G},
\end{equation}
which is the local physical-process first law for the selected
horizon bundle. For classical matter satisfying the null energy
condition, the same area response obeys
\(\delta A\geq0\), and hence
\(\delta S_{\rm hor}\geq0\). This local second-law statement compares
corresponding final cuts of the perturbed and reference bundles; it
does not imply that the perturbed horizon ceases to contract toward
the future tip.

The equilibrium geometry completes the horizon-law structure. The
conformal surface gravity has the uniform magnitude
\(\kappa=2/\alpha\) on the regular portions of both horizon branches,
providing the corresponding zeroth law. Moreover, every finite causal
diamond has strictly positive temperature. The limit \(T_D\to0\)
requires \(\alpha\to\infty\), for which the waist and tips recede to
infinity and the family of diamonds approaches complete Minkowski
spacetime,
\begin{equation}
\lim_{\alpha\to\infty}D_\alpha
=
\mathbb R^{2,1}.
\end{equation}
Zero temperature is therefore approached only in the limit in which
the finite causal horizon itself disappears, yielding the geometric
third-law statement for the finite-diamond family.

The higher-dimensional extension shows that, for spherical causal
diamonds in \(d\geq3\), the dimensional dependence is confined to the
transverse geometry and its measure. The equilibrium data and the
local mechanical laws consequently retain the same physical form as
in the explicit \((2+1)\)-dimensional construction.

Taken together, these results show that the equilibrium thermodynamics
and local mechanics of a causal-diamond horizon are not independent
structures. The same normalized conformal flow that fixes the
equilibrium temperature also supplies, in its near-horizon limit, the
Killing-energy flux entering the local physical-process law, while
the entropy response follows geometrically from the Raychaudhuri
evolution. Causal diamonds therefore support a coherent set of
horizon laws entirely within a finite region of Minkowski spacetime.
In this sense, the mechanical and thermodynamic structure usually
associated with horizons is not restricted to black-hole event
horizons, but can emerge from the causal organization of spacetime
itself.

\begin{acknowledgments}
We would like to thank Matheus G. Barbosa and Daniel Vanzella for reading a draft of this paper and for valuable feedback. We also thank Nada Eissa for insightful discussions during the early stages of this project. This work was partially supported by the Army Research Office (ARO) under Grant No.~W911NF-23-1-0202. G. V.-M. gratefully acknowledges the Center for Mexican American and Latino/a Studies at the University of Houston for its generous support through a Lydia Mendoza Fellowship.
\end{acknowledgments}

\section*{Data availability}
There are no publicly available research data or software supporting this manuscript. Requests for further information or data should be sent to the authors.

\appendix*
\renewcommand{\thesection}{} 
\makeatletter\def\theequation@prefix{A}\makeatother

\section{Higher-dimensional extension}
\label{app:higher-dimensional-extension}

The equilibrium and mechanical constructions in
Secs.~\ref{sec:equilibrium-thermodynamics} and
\ref{sec:mechanical-laws} were developed explicitly in \((2+1)\)
spacetime dimensions. For a spherical diamond in \(d\geq3\),
the radial-temporal geometry is unchanged, while the dimensional
dependence is confined to the \((d-2)\)-dimensional transverse geometry
and its measure. This Appendix records the replacements required to
reproduce the higher-dimensional results without repeating the
derivations of the main text.

The waist is now \(\Sigma_0\simeq S^{d-2}\), with unit-sphere metric
\(\gamma_{mn}(y)\). We denote its induced metric in the reference
geometry by
\begin{equation}
\begin{aligned}
q^{\rm ref}_{mn}(0;y)&=\alpha^2\gamma_{mn}(y),\\
A_0&=\int_{\Sigma_0}d^{d-2}y\,\sqrt{q_{\rm ref}(0;y)},\\
q_{\rm ref}(0;y)&\equiv\det q^{\rm ref}_{mn}(0;y).
\end{aligned}
\label{eq:app-d-dimensional-waist-data}
\end{equation}
The leading near-horizon geometry around the waist differs from
Eq.~\eqref{eq:CD-lorentzian-rindler-near-waist} only through this
transverse metric:
\begin{equation}
ds^2_{(\mathrm{NH})}
\simeq
-\kappa^2\ell^2d\eta^2
+
d\ell^2
+
q^{\rm ref}_{mn}(0;\,y)\,dy^m dy^n,
\qquad
\kappa
=
\frac{2}{\alpha}.
\label{eq:app-d-dimensional-near-waist}
\end{equation}
The Euclidean normal plane and its regularity condition are therefore
identical to those in the main text, so the Euclidean period and
temperature remain unchanged. Moreover, the regular Euclidean
continuation of the diamond geometry remains flat in the bulk, so the
complete regular bulk term vanishes. Repeating the regulated
Gibbons-Hawking-York calculation with the transverse integration in
Eq.~\eqref{eq:app-d-dimensional-waist-data} gives
\(I_D^{E}=-A_0/(4G)\)
\cite{York1972,GibbonsHawking1977,Laflamme1987}. As in the main text,
the absence of a mass term associated with the diamond geometry
implies \(E_D=0\), so the thermodynamic relation
reduces to \(I_D^{E}=-S_D\).
Collecting these results gives
\begin{equation}
\begin{aligned}
\beta_D&=\frac{2\pi}{\kappa}=\pi\alpha,\quad
T_D=\frac1{\pi\alpha},\\
I_D^E&=-\frac{A_0}{4G},\quad E_D=0,\quad S_D=\frac{A_0}{4G}.
\end{aligned}
\label{eq:app-d-dimensional-equilibrium-data}
\end{equation}
Thus, the equilibrium relations are unchanged; only the value of the
codimension-two waist area depends on the spacetime dimension.

For the mechanical response, let the angular labels of the selected
bundle lie in \(\mathcal U\subset S^{d-2}\), and define
\(d\Omega_{d-2}=\sqrt{\gamma(y)}\,d^{d-2}y\), with
\(\gamma(y)=\det\gamma_{mn}(y)\). The affine embedding
\(t=\lambda\), \(r=\alpha-\lambda\), and the null tangent
\(k=\partial_t-\partial_r\) are unchanged. At fixed \(\lambda\),
the reference cut has
\begin{equation}
\begin{aligned}
q^{\rm ref}_{mn}(\lambda;y)
&=(\alpha-\lambda)^2\gamma_{mn}(y),\\
\sqrt{q_{\rm ref}(\lambda;y)}
&=(\alpha-\lambda)^{d-2}\sqrt{\gamma(y)},\\
dA_{\rm ref}(\lambda)
&=(\alpha-\lambda)^{d-2}d\Omega_{d-2}.
\end{aligned}
\label{eq:app-d-dimensional-reference-cut}
\end{equation}
where
\(q_{\rm ref}(\lambda;\,y)
\equiv
\det q^{\rm ref}_{mn}(\lambda;\,y)\).
We impose the same near-waist restriction as in the main text, for
which
\begin{equation}
dA_{\rm ref}(\lambda)
\simeq
dA_{\rm ref}(0)
=
\alpha^{d-2}d\Omega_{d-2}.
\label{eq:app-d-dimensional-waist-measure}
\end{equation}
The local boost generator satisfies
\begin{equation}
\left.\xi^{(\mathrm{NH})\,\mu}\right|_{\mathcal H^+}
=
\kappa\lambda k^\mu,
\end{equation}
and the Killing-energy flux is
\begin{equation}
\delta E
=
\kappa
\int_{\mathcal U}dA_{\rm ref}(0)
\int_0^{\lambda_f}d\lambda\,
\lambda\,\delta T_{kk}.
\label{eq:app-d-dimensional-near-waist-flux}
\end{equation}

The reference expansion follows directly from
Eq.~\eqref{eq:app-d-dimensional-reference-cut}:
\begin{equation}
\theta_{\rm ref}(\lambda)
=
\frac{1}{\sqrt{q_{\rm ref}(\lambda;\,y)}}
\frac{d}{d\lambda}
\sqrt{q_{\rm ref}(\lambda;\,y)}
=
-\frac{d-2}{\alpha-\lambda}.
\label{eq:app-d-dimensional-reference-expansion}
\end{equation}
In \(d\) spacetime dimensions, the null Raychaudhuri equation contains
the focusing term \(-\theta^2/(d-2)\)
\cite{Raychaudhuri1955,Wald1984}. The spherical reference congruence
is shear-free, the shear contribution is quadratic and therefore
absent at linear order, and the twist vanishes because the generators
form a null hypersurface. Upon linearization, the factor \(d-2\) in
Eq.~\eqref{eq:app-d-dimensional-reference-expansion} cancels the
denominator in the focusing term. Consequently, the response obeys
exactly the same equation as in the main text:
\begin{align}
\frac{d\,\delta\theta}{d\lambda}
&=
\frac{2}{\alpha-\lambda}\,\delta\theta
-
8\pi G\,\delta T_{kk},
\label{eq:app-d-dimensional-linearized-raychaudhuri}
\\
\delta\theta(\lambda;\,y)
&=
\frac{8\pi G}{(\alpha-\lambda)^2}
\int_\lambda^{\lambda_f}d\lambda'\,
(\alpha-\lambda')^2
\delta T_{kk}(\lambda';\,y),
\label{eq:app-d-dimensional-expansion-response}
\end{align}
where the second line uses the same final matching condition
\(\delta\theta(\lambda_f;\,y)=0\) as
Eq.~\eqref{eq:mechanical-final-expansion-matching}.

The higher-dimensional area response is obtained by replacing the
transverse length density of the main text by \(\sqrt q\). Linearizing
the transverse measure gives
\begin{equation}
\frac{d}{d\lambda}
\left[
\frac{\delta(\sqrt q)(\lambda;\,y)}
{\sqrt{q_{\rm ref}(\lambda;\,y)}}
\right]
=
\delta\theta(\lambda;\,y),
\qquad
\delta(\sqrt q)(0;\,y)
=
0.
\label{eq:app-d-dimensional-fractional-area}
\end{equation}
Substitution of
Eq.~\eqref{eq:app-d-dimensional-expansion-response} and the same
exchange of integration order used in the derivation of
Eq.~\eqref{eq:mechanical-endpoint-area-response-exact} give
\begin{equation}
\delta A
=
\frac{8\pi G(\alpha-\lambda_f)^{d-2}}{\alpha}
\int_{\mathcal U}d\Omega_{d-2}
\int_0^{\lambda_f}d\lambda\,
\lambda(\alpha-\lambda)\delta T_{kk}.
\label{eq:app-d-dimensional-exact-area-response}
\end{equation}
For fixed \(d\), the leading near-waist area response is
\begin{equation}
\delta A
\simeq
8\pi G
\int_{\mathcal U}dA_{\rm ref}(0)
\int_0^{\lambda_f}d\lambda\,
\lambda\,\delta T_{kk}.
\label{eq:app-d-dimensional-near-waist-area}
\end{equation}

Therefore, the dimensional dependence cancels from the local horizon
laws:
\begin{equation}
\delta E
\simeq
\frac{\kappa}{8\pi G}\delta A
=
T_D\,\delta S_{\rm hor},
\qquad
\delta S_{\rm hor}
\equiv
\frac{\delta A}{4G}.
\label{eq:app-d-dimensional-first-law}
\end{equation}
The kernel in
Eq.~\eqref{eq:app-d-dimensional-exact-area-response} is non-negative
for \(0\leq\lambda\leq\lambda_f<\alpha\). Hence, for classical
matter satisfying the null energy condition,
\begin{equation}
\delta T_{kk}\geq0
\quad\Longrightarrow\quad
\delta A\geq0,
\qquad
\delta S_{\rm hor}\geq0.
\label{eq:app-d-dimensional-second-law}
\end{equation}
Finally, because \(\xi\) and \(\kappa=2/\alpha\) are
independent of the angular coordinates, the zeroth law and the absence
of a finite zero-temperature diamond are unchanged. The
higher-dimensional dependence is therefore confined to the transverse
measure and the value of \(A_0\); the equilibrium data and the local
mechanical laws retain the forms derived in the main text.


\end{document}